\documentclass[aps,twocolumn,showpacs]{revtex4}

\usepackage{epsf}

\begin{document}

\begin{widetext}

\title{Space Charge Expansion for Time-resolved
Spin-Polarized Electron Spectroscopy}%

\author{K.~Aulenbacher$^{1}$,  A.~V.~Subashiev$^{2}$, V.~Tioukine$^{1}$, \\
D.~Bender$^{1}$,  and  Yu.~P.~Yashin$^{2}$}

\affiliation{{$^{1}$~Institut f\"ur Kernphysik der Univ. Mainz, J.J. Becherweg 45,
D-55099 Mainz, Germany} \\{$^{2}$~Department of Experimental Physics, State
Polytechnical University, Polytechnicheskaya 29, St~Petersburg, 195251, Russia}}

\email{arsen@subashiev.hop.stu.neva.ru}
\begin{abstract}
Time resolved spin-polarized electron photoemission spectra are investigated as a
function of excitation pulse energy for the heterostructures with a single strained
layer and with a strained-well superlattice. At an average current exceeding 10 nA the
emission pulse profiles are modified by the space charge pulse expansion during the
electron transport to detector. The pulse expansion enables the separation of
electrons that have spent minimum time in the sample. For the superlattice structure
these electrons showed maximum polarization above 90 \%. Variation in the pulse
profiles for the two structures is interpreted as resulting from the difference in the
effective NEA values.

\begin{center}
\end{center}

\end{abstract}
\maketitle
\end{widetext}

\tolerance=7000

\section*{Introduction}%
 Ultrafast spin-sensitive spectroscopy has drawn much attention in the last decade
due to the proposal of novel electronic devices and circuits [1]. Spin-polarized
spectroscopy of photoelectrons emitted in vacuum is shown to be effective in the
studies of the absolute values of polarization and its temporal variations [2],
polarized electron energy distribution [3], and the polarized electron kinetics at the
surface [4].

Availability of high-intensity subpicosecond lasers has prompted
the development of electron beam photocathodes
with high emission currents. A high beam current leads to
the space-charge effects and temporal expansion of
the photoemission pulse [5]. In this report we use
the pulse temporal expansion to improve time resolution in the
spin-resolved photoemission studies of the semiconductor
photocathode structures, known to be very effective as highly
polarized electron sources.

The time resolved polarization measurements combined with space
charge pulse expansion enabled us to distinguish polarization
losses at the subsequent photoemission stages and to filter out
the electrons that have spent minimum time in the sample and have
maximum possible polarization.
\section*{Experimental details}
 Two heterostructures with deformation-splitted valence
band were investigated: (1) MOVPE grown GaAs$_{0.95}$P$_{0.05}$ (120 nm) /
GaAs$_{0.68}$ P$ _{(0.32}$ (0.5 mkm) Strained Layer Heterostructure (SLH) and (2) MBE
grown modulation-doped In$_{0.16}$Al$_{0.2}$Ga$_{0.64}$As / Al$_{0.28}$Ga$_{0.72}$As
Strained-well Superlattice (SWSL) structure consisting of 12 pairs of 5 mn$\times$ 4
nm layers grown on Al$_{0.3}$Ga$_{0.7}$As buffer layer, with narrow-band 6 nm- thick
GaAs overlayer, both fabricated at Ioffe Physico-Technical Institiuts, St Petersburg,
Russia.

Mode-locked Ti:AlO laser  with the wavelength $\lambda$= 799 nm close to polarization
maximum in the emitted current was used to generate the emission pulses. The
excitation pulse power was varied to result in variation of the average photoemission
current in the range of $I_{\rm emi}$=  1 -- 300 nA.

Time-resolved polarization was obtained by the measurements of spin polarization of
the electrons accelerated to the energy of 100 keV and passed through a microwave
deflection cavity [6]. The apparative time resolution determined mainly by the finite
beam size on the analyzer slit was estimated to be $\tau_{app}$ = 2.5$\pm$ 0.5 ps.
\section*{Results and discussion}
Polarization spectra measured at cw regime were typical for these structures and
showed polarization maxima centered  at $\lambda = 810$ nm for the SLH with $P =$ 83.8
\% at quantum efficiency $QE = 0.3$ \%  and at $\lambda = 799$ for the SWSL with $P =$
84.4 \% and $QE = 0.7$ \% , which is very close to the former results for these
structures.

The emission current profiles for the SLH sample show the growth of the pulse width
with $I_{\rm emi}$ due the space charge expansion, the rise time and decay time of the
pulses remaining constant within the range of 3 ps for all current values. The
observed variation of the polarization along the pulse at all $I_{\rm emi}$ is in the
range of 2 \%, while the variations of polarization at the pulse front and back are in
the range of $\approx$ 5 \%.

The pulse shapes for the SWSL structure show close to triangular shape of the emission
pulse at low  $I_{\rm emi}$ and a specific variation of the profiles with the pulse
energy accompanied by a sizable temporal dispersion of the spin polarization along the
pulse.

\begin{figure}[b] \leavevmode
\centering{\epsfxsize=7.9cm,\epsfysize=6.5cm,\epsfbox{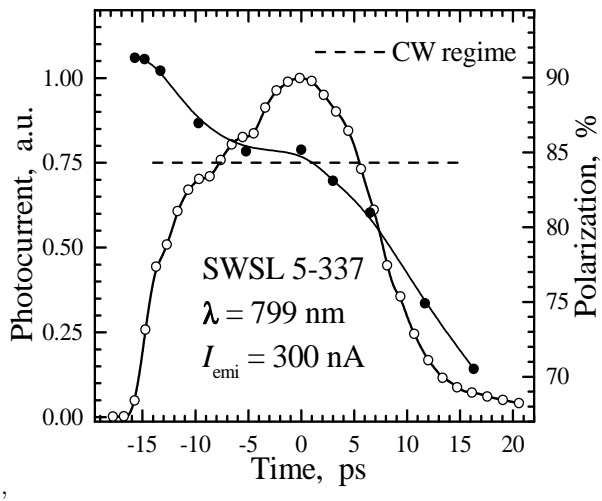}}
\caption[]{Polarization distribution along the emission pulse profile.}
\end{figure}

The temporal distribution of polarization and the photoemission current profile at the
maximum value of the average photocurrent $I_{\rm emi}$ = 300 nA are presented in Fig.
1. The maximum value of polarization at the front of the emission pulse grows with the
pulse charge and achieves $P \approx$ 91 \% at $I =$ 300 nA. The pulse length as a
function of the excitation energy is shown for the two structures in Fig. 2. For both
structures the length of the pulse, measured as a time interval, during which 90 \% of
the pulse charge is emitted, grows starting from about 10 pA current and then closely
follows a square-root dependence, $t_{90} \propto \sqrt{Q} $ where $Q$ is the pulse
charge.

The profile of the current pulse is determined by the initial distribution of the
electron velocities and emission moments. At small currents the average pulse duration
depends on the extraction time from the working layer  $t_{\rm d}$ and the
acceleration time dispersion $\tau_{\rm acc}$ due to the spread of the electron
initial velocities normal to the surface in the samples with high Negative Electron
Affinity (NEA).

For the analysis of the experimental profiles we have used as an initial distribution
function, a distribution generated by the diffusion of electrons from the working
layer [2] and their subsequent energy relaxation in the band bending region [3]. The
calculated pulse shapes differ for the case of $\tau_{\rm acc} > t_{\rm d}$ (large
NEA) and for the opposite case. The profile for the small values of NEA (SLH) is
determined by the experimental resolution and is close to Gaussian. For higher NEA the
shape is non-gaussian and is close to experimental (triangular) assuming the
homogeneous distribution of the electron velocities.

The electric charge of the pulse generates the electric field which changes the
acceleration conditions though the field of the pulse remains by two orders of
magnitude lower than the external accelerating field. The dispersion of the arrival
time caused by the Coulomb repulsion $ \delta t_0 = t_0 {{\delta a}/ a}, $ where
$\delta a ={ eQ / {S \epsilon_O m}}$, $S$ is the excitation spot area, and ${\delta a
/ a} \approx 10^{-2}$ for the pulse charge of 4.5 FC. The estimation of the spread
time due to Coulomb repulsion gives experimentally observed values and results in
close to a square root dependence of the pulse length on its charge.
\begin{figure}[h]
\leavevmode \centering{\epsfxsize=7.8cm,\epsfysize=6.4cm,\epsfbox{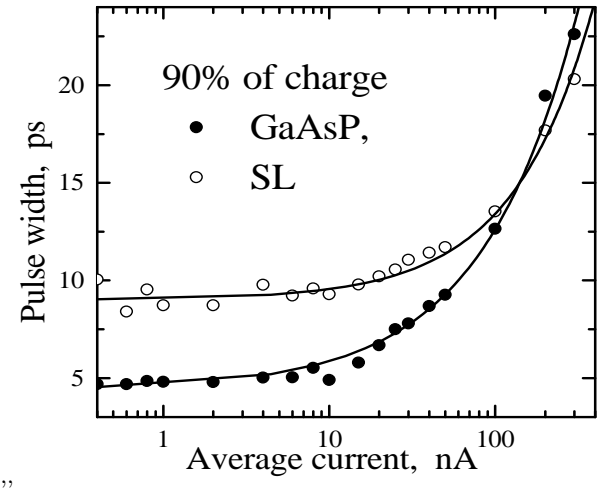}}
\caption[]{Emission pulse length for strained GaAsP layer and for SW SL structures for
different values of the $I_{\rm emi}$.}
\end{figure}
Note that the time of electron flight to the registration region and the time of
electron spread resulting from the acceleration time dispersion and Coulomb repulsion
are considerably larger than the pulse duration itself. Therefore, after the first few
picoseconds the dispersion of the initial velocity becomes unimportant and the
emission pulse is only stretched by the repulsion during the acceleration.

In photoemission from thin layers two competing spin depolarization processes are the
electron spin relaxation in extraction to the surface and the spin relaxation during
the electron energy relaxation in the Band Bending Region (BBR). Separation of a very
fast process of the electron relaxation in the BBR (having sub-picosecond time scale)
from a more slow process of the electron extraction from the working layer (with
typical times of 1-2 ps, depending on the thickness of the structure) at low $I_{\rm
emi}$ is not possible.

At the large $I_{\rm emi}$  values  in case of $\tau_{\rm acc} <
 t_{\rm d}$,  (SLH sample)
the electrons in the expanded pulse have experienced the energy relaxation process and
differ only in time, spent in the working layer. Minimal variation of polarization
along the plateau (less than 2 \%) reflects minimal polarization losses in this
process, in agreement with the former conclusions [2] based
on the dependence of the polarization on the working layer thickness.

The front of the pulse is associated with electrons having a small velocity spread
near its maximum value and reduced time, spent in the layer. Thus, the polarization
losses in the energy relaxation in the BBR can be estimated. Equal polarization
decline at the front and the final parts of the pulse should be entirely due to
depolarization in the energy relaxation process in the BBR. The obtained value (close
to 4 \%) is in line with estimation of Ref. [7].

For $\tau_{\rm acc} > t_{\rm d}$ (SWSL sample) the front part of the pulse is also
associated with high-velocity electrons extracted close to the excitation moment. Then
the polarization decay during the electron relaxation time in BBR result in average
polarization decay of  6 \% in the emission pulse. Note, that in this case the
polarization losses in transport remain constant along the final part of the pulse,
while the dispersion of polarization losses in BBR should prevail.
\section*{Conclusion}
The achieved time resolution, increased by the space charge pulse expansion, allowed
us to separate the first high-velocity electrons emitted at the high end of the
electron energy distribution close to the the excitation moment that have not
experienced spin relaxation during the processes of transport to the surface or energy
relaxation at the surface potential well. Their polarization is found to be above 90
\%, which is apparently the highest spin polarization value registered for the
present.

{\em Acknowledgements}

This work was supported by CRDF under grant RP1-2345-ST-02, and by NATO under grant
PST.CLG 979966.

\end{document}